# Peer-to-Peer Cloud Provisioning: Service Discovery and Load-Balancing

*Rajiv Ranjan, Liang Zhao, Xiaomin Wu, and Anna Liu*
*SA Project, CRC Smart Services*
*Service Oriented Computing Research Group,*
*School of Computer Science and Engineering*
*University of New South Wales, Australia*
Email: *{rajivr, lzhao, xmwu432, annaliu}@cse.unsw.edu.au*

## ABSTRACT

Clouds have evolved as the next generation platform that facilitates creation of wide-area on-demand renting of computing or storage services for hosting application services that experience highly variable workloads and requires high availability and performance. Inter-connecting Cloud computing system components (servers, VMs, application services) through peer-to-peer routing and information dissemination structure is essential to avoid the problems of provisioning efficiency bottleneck and single point of failure that are predominantly associated with traditional centralized or hierarchical approaches. These limitations can be overcome by connecting Cloud system components using a structured peer-to-peer network model (such as Distributed Hash Tables (DHTs)). DHTs offer deterministic information/query routing and discovery with close to logarithmic bounds with regards to network message complexity. By maintaining a small routing state of $O(\log n)$ per VM, a DHT structure guarantees deterministic look ups in a completely decentralized and distributed manner.

This chapter presents: (i) a layered peer-to-peer Cloud provisioning architecture; (ii) a summary of the current state-of-the-art in Cloud provisioning with particular emphasis on service discovery and load-balancing; (iii) a classification of the existing peer-to-peer network management model with focus on extending the DHTs for indexing and managing complex provisioning information; and (iv) the design and implementation of novel, extensible software fabric (*Cloud peer*) that combines public/private clouds, overlay networking and structured peer-to-peer indexing techniques for supporting scalable and self-managing service discovery and load-balancing in Cloud computing environments. Finally, an experimental evaluation is presented that demonstrates the feasibility of building next generation Cloud provisioning systems based on peer-to-peer network management and information dissemination models. The experimental test-bed has been deployed on a public cloud computing platform, Amazon EC2, which demonstrates the effectiveness of the proposed peer-to-peer Cloud provisioning software fabric.

## 1 INTRODUCTION

Cloud Computing [1][2][3] has emerged as the next generation platform for hosting business and scientific applications. It offers infrastructure, platform, and software as services that are made available as on-demand and subscription-based services in a pay-as-you-go model to users. These services in Information and Communication Technology (ICT) industry are respectively referred to as Infrastructure as a Service (IaaS), Platform as a Service (PaaS), and Software as a Service (SaaS). Adoption of Cloud computing platforms [5][6][7][13][14][18] as an application provisioning environment has the following critical benefits: (i) software enterprises and startups with innovative ideas for new Internet services are no longer required to make large capital outlays in the hardware and software infrastructures to deploy their services or human expense to operate it; (ii) government agencies and financial organisations can use Cloud services as an effective means for cost cutting by leasing their IT services hosting and maintenance responsibility to external cloud(s); (iii) organisations can more cost effectively manage peak-load by using the cloud, rather than planning and building for peak load, and having under-utilised servers sitting there idle during off peak time, and (iv) failures due to natural disasters or regular system maintenance/outage may be managed more gracefully as services may be more transparently managed and migrated to other available cloud resources, hence enabling improved service level agreement (SLA).

*The process of deploying application services on publically accessible clouds (such as Amazon EC2 [14] ) that expose their capabilities as a network of virtualized services (hardware, storage, database) is known as Cloud Provisioning.* The Cloud provisioning process consists of two key steps [37]: (i) *VM Provisioning*, this involves instantiation of one or more VMs on physical servers hosted within public or private Cloud computing environments. The selection physical server for hosting VMs in a cloud is based on number of mapping requirements including available memory, storage space, and proximity of the parent cloud; and (ii) *Application Service Provisioning*, the second step is mapping and scheduling of requests to the services that are hosted within a VM on a set of VMs. In this chapter, we mainly focus on the second step, where given a set of VMs that are hosting different types of application services, how to dynamically distribute the incoming requests among the services in a load-balanced and decentralized manner.

Cloud provisioning from a business services point of view involves deriving cloud-based application component deployments driven by expected performance (QoS). Clouds offer unprecedented pool of software and hardware resources, which gives businesses a unique ability to handle the temporal variation in their service demands through dynamic provisioning or de-provisioning of capabilities. Actual usage patterns of many enterprise services (business applications) vary over time, most of the time in an unpredictable way. Whenever there is a variation in temporal and spatial locality of workload such as number of concurrent users, total users, and load conditions; each application component must dynamically scale (application service elasticity) to offer good quality of experience to users, and maintain an optimal usage of cloud resources. Cloud-enabling any class of application service would require developing models for service placement, computation, communication, storage with emphasis on important scalability requirements.

Services provisioned across multiple clouds are located in different network domains that may use heterogeneous addressing and naming schemes (public addresses, private addresses with NAT, etc.). In general, services would require all their distributed components to follow a uniform IP addressing scheme (for example, to be located on the same local network), so it becomes mandatory to build some kind of overlay network on top of the physical routing network that aids the service components in undertaking seamless and robust communication. Existing implementation including VPN-Cubed, OpenVPN [19] provides an overlay network that allows application developer to control addressing, topology, protocols, and encrypted communications for services deployed across multiple clouds (private and public) sites. However, these implementations do not provide capabilities related to decentralized service discovery, monitoring, and load-balancing across VM instances.

Currently, one of the prominent Cloud service providers Amazon EC2 offers two services namely CloudWatch [8] and Elastic Load Balancer [9]. CloudWatch is a web service that is responsible for monitoring Amazon Web Service (AWS) cloud resources such as Amazon EC2. It provides application developers with the important details related to a VM instance's resource utilization, operational performance, disk reads and writes, and network information. Developers are required to attach EC2 instances that they would like to monitor to the centralized CloudWatch service. Fundamentally, CloudWatch is a centralized monitoring service that can be associated with numerous EC2 instances deployed over multiple Amazon Cloud sites or Zones. Similarly, Amazon's Elastic Load Balancer (ELB) is also a centralized web service. The role of the ELB is to automatically distribute incoming application across EC2 instances. The load-balancer can control request load-balancing across single Cloud site as well as multiple cloud sites. The load-balancer performs provisioning related decision based on dynamic monitoring data reported by the CloudWatch service. In line with CloudWatch, Microsoft Azure platform also has a centralized service called Azure Fabric Controller (FC) [13], which monitors, maintains and provisions machines to host the applications that the developer creates and deploys in the Microsoft Cloud.

However, centralized approaches to service discovery [20], monitoring, and load-balancing have several critical design limitations including: (i) single point of failure; (ii) lack scalability; (iii) high network communication cost (such as network bottleneck, congestion) at links leading to the service; (iv) requirement of high computational power (which may be not feasible with commodity machines that public clouds offer) to serve a large number of resource look-up and updated queries on the server running the central service. Additionally, recent studies conducted in Grid Computing research verified that centralized monitoring and discovery services [29] such as R-GMA, MDS, and Hawkeye fail to scale beyond *300* concurrent users i.e. the throughput declines below acceptable levels.

As Clouds become ready for mainstream acceptance, scalability [10] of services will come under more severe scrutiny since at that time Cloud providers will have to support an increasing number of online services, each being accessed by massive numbers of global users round the clock. To overcome the aforementioned limitations, fundamental Cloud services for discovery, monitoring, and load-balancing should be decentralized by nature and different service components (VM instances, application elements) must interact to adaptively maintain and achieve the desired system wide connectivity and behavior.

The rest of this chapter is organized as follows: First, the benefits related to provisioning of applications on Clouds are discussed. Next, a brief discussion on the algorithmic and system design challenges related to Cloud provisioning. Then, a layered approach to architecting peer-to-peer Cloud provisioning system is presented. This is followed by some survey results on Cloud provisioning capabilities in leading commercial public clouds. The finer details related to architecting peer-to-peer Cloud service discovery and load-balancing techniques over DHT overlay is then presented, followed by a discussion of the design and implementation of peer-to-peer Cloud provisioning (Cloud peer) software fabric. Lastly, we present the analysis and experimental results of the peer-to-peer Cloud provisioning implementation across a public Cloud (Amazon EC2) environment.

## 2 BENEFITS OF PROVISIONING APPLICATIONS ON CLOUDS

There are significant business and infrastructure level benefits [1][12][24] of using clouds as an application hosting platform. These include:

- On-demand dynamic computing: Service users can dynamically provision or de provision computing and storage resources from the existing Cloud infrastructures driven by business demands. These services are often presented as virtualized instances (VMWare, Microsoft Hypervisor, or Xen) that act as the hosting and execution environment for application components (such as business processes). Companies (particularly start-ups) with innovative ideas for new Internet services are no longer required to make large capital outlays for the hardware and software infrastructure on which to deploy their services, or to pay for the human resources for operating it. Furthermore, the enterprises which traditionally provision their services on private clouds can handle the peak demands by dynamically leasing capabilities from public clouds. Also, adapting to downtime including failures like natural disasters and regular system maintenance, can be potentially more graceful as services may be transparently migrated to public clouds thus resulting in better continuity of operations and potentially achieving better Service Levels.

- Zero upfront investment and Pay per use: Through leveraging Cloud computing platforms, customers (government departments, SaaS providers) are able to save large initial expenditures related to setting up and administering the basic infrastructure. These expenditures are related to real-estate, hardware (racks, machines, routers, backup power supplies), hardware management (cooling, power supply), and operations personnel. By deploying applications on clouds, customers are freed from the burden of infrastructure planning and capacity management. Cloud-based hosting solutions present low financial risk because system size can grow as the demand for services picks up. Since clouds offer services under a usage-driven model (pay per use) customers pay for only what they actually use.

- Proximity aware server pooling: Performance of Internet-based applications such as online-gaming, virtual reality and social networking depend on the network proximity of services to their users. However, it is typically not possible for an application service provider to establish data centers at all possible locations to counter the network delays to users in all locations. As a result, Internet service providers may not be able to meet service satisfaction level for all of their users. With the advent of Cloud computing and the significant investment in data centers at multiple geographical locations from Cloud vendors, SaaS providers [18] can selectively pool the servers at locations that better suit the service users' needs. Internet-scale service providers such as Facebook and Myspace can dynamically improve their service quality by optimizing the service placement according to users' geographical proximity and Quality of Service (QoS) needs.

- Ubiquitous network access: Clouds offer their capabilities over a unified network, which can be accessed through standard mechanisms that are language and platform independent (such as web service interfaces). Such standard mechanisms can transparently support the use of Cloud services from a variety of user platforms including mobile phones, laptops and PDAs.

## 3 THE PROBLEM: MANAGING COMPLEXITY OF CLOUD PROVISIONING

Cloud infrastructures are heterogeneous, large-scale, highly dynamic and geographically distributed. Similarly, application services coming in from a large set of public users, brings multiple, independent, and highly distributed software elements embodied in services. Further, the application services arriving at a public cloud also have radically different application characteristics and workload profiles, ranging from the traditional e-Commerce application types, to the newer social networking and collaboration applications, to the enterprise business applications such as CRM and ERP, and to the computing and data intensive type of applications. To appropriately respond to the aforementioned complexity and challenges, application provisioning technologies and approaches should adapt to changing application states and behaviors [4] (leave, join, failure, utilization, availability, workload patterns) of the Cloud computing environments in accordance with high-level guidance specified by the Cloud application developers. Fig.1 shows a reference diagram explaining different dynamics and complexities ambient in emerging Cloud computing environments.

For many enterprises, there is a large amount of IT assets in house, in the form of line of business applications that are unlikely to ever be migrated to the cloud. This may be due to the fact that sensitive data resides in the application and hence cannot reside in a public cloud due to privacy and security issues. Also, increasingly, customer facing departments are creating more web applications to serve customers, which rely on some Cloud platforms, or are services themselves that are deployed on a cloud. As a result, there is a need to look into the scenario and issues related to integration and interoperability between the software on premises and services in the cloud.

To meet these requirements, next generation Cloud provisioning techniques and services should be able to: (i) dynamically adapt to performance needs, failure, and leave and join of hardware and software, including VMs, servers, storage, software, applications, and networks; (ii) discover and monitor state of services in completely decentralized and distributed manner; (iii) accomplish coordinated and load-balanced provisioning of VMs and application services; (iv) handle spikes in service demand (workload) through dynamic scaling in of services from other public clouds; and (v) handle authentication and authorization of services for users; provisioning users' access; federated security model.

A fundamental challenge in managing the Cloud provisioning system is to maintain a consistent connectivity between the components (self-organization) (Parashar & Hariri, 2007). This challenge cannot be overtaken by introducing a central network model to connect the components, since the information needed for managing the connectivity and making the decisions is completely decentralized and distributed. Further, centralized network model [29] does not scale well, lacks fault-tolerance, and requires expensive server hardware infrastructure. System components can *leave (VM instance destruction), join (VM creation), and fail (service outage)* in a dynamic fashion; hence it is an impossible task to manage such a network centrally. Therefore, an efficient decentralized or peer-to-peer solution is mandatory that can gracefully adapt, and scale to the changing conditions.

In peer-to-peer organization of Cloud provisioning systems [16] both control and decisions making are decentralized by nature and where different system components interact together to adaptively maintain and achieve a desired system wide behavior. A distributed Cloud provisioning configuration is considered to be decentralized "if none of the components in the system are more important than the others, in case that one of the component fails, then it is neither more nor less harmful to the system than caused by the failure of any other component in the system".

A possible way to efficiently interconnect the distributed system components can be based on structured peer-to-peer network models. In literature, structured peer-to-peer models are more commonly referred to as the DHTs [31]. DHTs provide hash table like functionality at Internet scale. DHTs such as Chord [15], CAN [25], Pastry [27], and Tapestry [31] are inherently self-organizing, fault-tolerant, and scalable. DHTs provide services that are light-weight and hence, do not require an expensive hardware platform for hosting, which is an important requirement as regards to building and managing Cloud provisioning systems that aggregate massive number of commodity servers and virtualized instances hosted within them. A DHT is a distributed data structure that associates a key with a data. Entries in a DHT are stored as a (key, data) pair. A data can be looked up within a logarithmic overlay routing hops if the corresponding key is known.

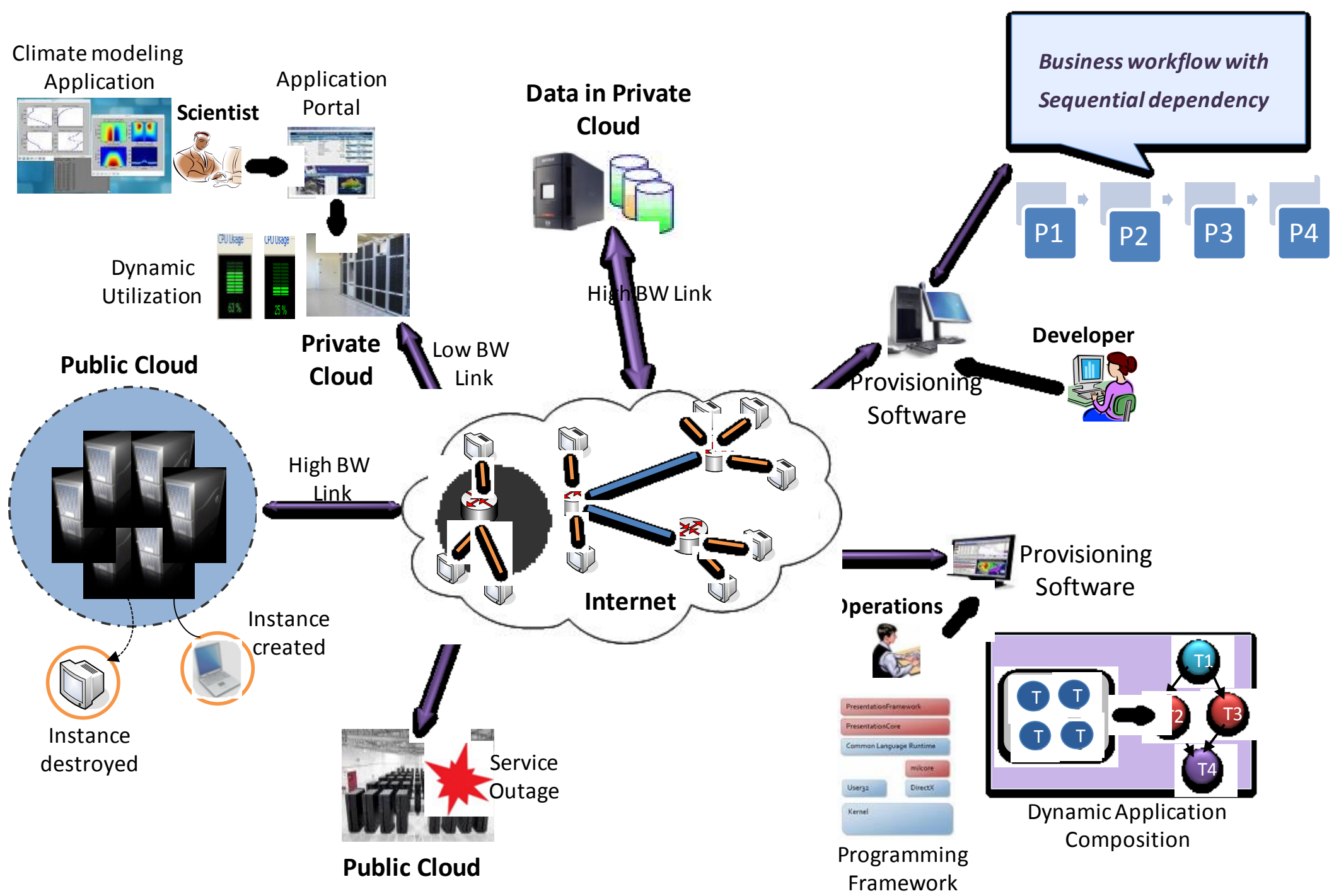

Figure 1: A diagram showing complexity in Cloud provisioning.

## 4 LAYERED PEER-TO-PEER CLOUD PROVISIONING ARCHITECTURE

This section presents information on various architectural elements that form the basis for peer-to-peer Cloud provisioning architecture. It also presents an overview of the applications that would benefit from the proposed architecture, which envisages a hosting infrastructure consisting of multiple geographically distributed private and public clouds owned by one or more service providers. Fig. 2 shows the layered design of the peer-to-peer Cloud provisioning architecture. Physical Cloud servers, along with core middleware capabilities, form the basis for delivering IaaS. The user-level middleware aims at providing PaaS capabilities. The top layer focuses on application services (SaaS) by making use of services provided by the lower layers. PaaS/SaaS services are often developed and provided by 3rd party service providers, who are different from IaaS providers.

### *4.1 Cloud Applications (SaaS)*

Popular Cloud applications include Business to Business (B2B) applications, traditional eCommerce type of applications, enterprise business applications such as CRM and ERP, social computing such as Facebook and MySpace, and compute, data intensive applications and Content Delivery Networks (CDNs). These applications have radically different application characteristics and workload profiles, and hence, to cope with the variation in temporal and spatial locality of service request, the application services must be supported by a Cloud provisioning infrastructure that dynamically scale the deployed services in order to achieve good performance, optimal resource usage, and hence offer quality experience to its end users.

### *4.2 Development Framework Layer*

This layer includes the software frameworks such as Web 2.0 Interfaces (Ajax, IBM Workplace, and Visual Studio.net Azure plug-in) that help developers in creating rich, cost-effective, user-interfaces for browser-based applications. The layer also provides the data intensive, parallel programming environments (such as MapReduce, Hadoop, Dryad) and composition tools that ease the creation, deployment, and execution of applications in Clouds.

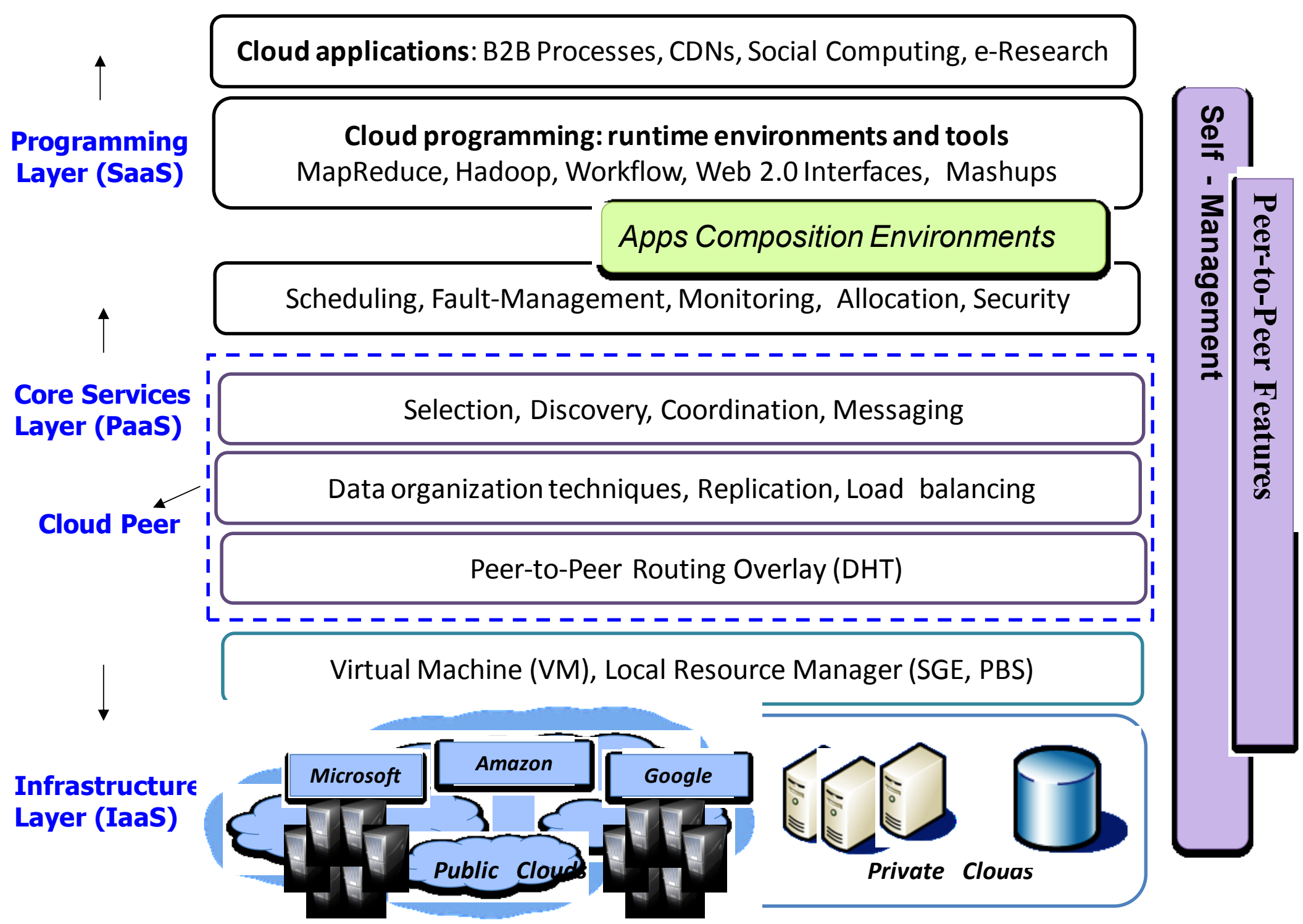

Figure 2: A layered peer-to-peer Cloud provisioning architecture.

### *4.3 Core Services Layer (PaaS)*

This layer implements the platform level services that provide runtime environment-enabling Cloud computing capabilities to application services built using User-Level Middlewares. Core services at this layer include Scheduling, Fault-Management,

Monitoring, Dynamic SLA Management, Accounting, Billing and Pricing. Further, the services at this layer must be able to provide support for decentralized coordinated interaction, scalable selection, and messaging between distributed Cloud components. Some of the existing services operating at this layer are Amazon EC2's CloudWatch and Load-balancer service, Google App Engine, Microsoft Azure's fabric controller, and Aneka [23].

To be able to provide support for decentralized service discovery [20] and load-balancing between cloud components (VM instances, application services); novel Distributed Hash Table (DHT)-based PaaS layer services, techniques, and algorithms need to be developed at this layer for supporting complex interactions with guarantees on dynamic management. In Fig. 2, this component of PaaS layer is shown as Cloud peer service. Architecting Cloud services based on decentralized network models or overlays (such as DHTs) is significant since DHTs are highly scalable, can gracefully adapt to the dynamic system expansion (new host/VM/service instantiation) or contraction (host/VM/service instance destruction) and outage, and are not susceptible to single point of failure in massive scale, inter-networked private and public cloud environments.

### *4.4 Infrastructure Layer (IaaS)*

The computing power in Cloud computing environments is supplied by a collection of data centers that are typically installed with many thousands of servers. At the IaaS layer there exists massive physical servers (storage servers and application servers) that power the data centers. These servers are transparently managed by the higher level virtualization services and toolkits that allow sharing of their capacity among virtual instances of servers. These virtual machines (VMs) are isolated from each other, which aids in achieving fault tolerant behavior and the isolation of security contexts.

Another trend in Cloud usage is combination of private clouds with public clouds, in order to attend unexpected or periodic peaks in local demand without investing in acquiring new equipment for the local infrastructure. Resources from the data center may be either available for public in general (public clouds) or may be restricted to users belonging to the organization that owns the data center (private clouds). It is also possible to have hybrid models, in which resources are leased from the public cloud whenever the private cloud cannot cope with the incoming demand.

## 5 CURRENT STATE-OF-THE-ARTS AND PRACTICE IN CLOUD PROVISIONING

Key players in public Cloud computing domain including Amazon, Microsoft, Google App Engine, Eucalyptus [41], and GoGrid [40] offer a variety of pre-packaged services for monitoring, managing and provisioning resources. Amazon EC2 offers three services namely Elastic Load Balancer, Auto Scaling and CloudWatch. Eucalyptus uses a hierarchical controller structure. Windows Azure provides a capability called the Azure Fabric Controller. On the other hand, GoGrid Cloud Hosting provides a service named F5 Load Balancer. While, Google App Engine supports their own proprietary auto-scaling technology, with which Google applications also uses behind the scene.

All leading Cloud computing platforms are able to perform load balanced provisioning and auto scaling to some degree. However, the techniques implemented in each of these Clouds vary: some use a centralized approach, some use a hierarchical approach. Some of these platforms apply distributed state replication of critical services for achieving fault-tolerant behavior. Some platforms can be managed automatically; some offer mixed management, i.e. both manual and automated; some enables complete automatic scaling behind the scene such as Google App Engine, trading off flexibility and architecture constraint; while some enforce developers to take on the responsibility do plan, design and implement scaling and resource allocation tasks manually.

The three Amazon Web Services (AWS), Elastic Load Balancer, Auto Scaling and CloudWatch together expose functionalities which are required for undertaking provisioning of application services on Amazon EC2. Elastic Load Balancer service automatically provisions incoming application workload across available Amazon EC2 instances. It also pays close attention on health conditions of instances and based on that it performs traffic rerouting from faulty instances to healthy ones. Auto Scaling service can be used to dynamically scale-in or scale-out the number of Amazon EC2 instances for handling changes in service demand patterns. And finally the CloudWatch service can be integrated with an application provisioner for collecting real-time information related to application services and VMs. The data monitored by CloudWatch service are required by other AWS services including Elastic Load Balancer and Auto Scaling services. Although AWS services are published and hosted separately, all functions of these services are supported via WSDL APIs, hence enabling the simple integration of services.

| **Cloud Platforms** | **Load Balancing** | **Provisioning** | **Auto Scaling** |
|---|---|---|---|
| Amazon Elastic Compute Cloud | √ | √ | √ |
| Eucalyptus | √ | √ | × |
| Microsoft Windows Azure | √ | √ (fixed templates so far) | √ (Manually at the moment) |
| Google App Engine | √ | √ | √ |
| GoGrid Cloud Hosting | √ | √ | √ (Programmatic way only) |

Table 1: Summary of provisioning capabilities exposed by public Cloud platforms.

Eucalyptus is an open source Cloud computing platform. The system is composed of three controllers that manage the virtualization environment based on centralized and hierarchical network and service structure. The three controllers are Node Controller, Cluster Controller and Cloud Controller, respectively being in charge of managing physical resources for virtual machines, coordinating Node Controllers within the same availability zone, processing connections from external clients and administrators. Among the three controllers, the Cluster Controller is a key component that undertakes activities related to application service provisioning and load balancing. Each Cluster Controller is hosted on the head node of a cluster to enable an availability zone, while inter-connecting outer public networks and inner private networks together. By monitoring the state information of instances in the pool of server controllers, the Cluster Controller can select the available service/server for provisioning incoming requests. However, as compared to AWS, Eucalyptus still lacks of some of the critical functionalities, such as auto scaling, live migration and support for built-in provisioner.

Azure Fabric Controller aims to be a highly redundant and fault-tolerant service designed for monitoring, maintaining and provisioning Cloud servers that hosts applications.

Fundamentally, Windows Azure Fabric has a weave-like structure, which is composed of servers, load balancers, and edges (power, Ethernet and serial communications). The Fabric Controller manages the servers in the Windows Azure Fabric differently depending on various factors such as service types. If a server is a hardware load balancer then it is managed through a custom driver interface which is implemented from an Azure supported driver model for compatibility purpose. If a Cloud is marked as a service node, then a built-in service, named Azure Fabric Controller Agent that runs in the background and tracks the current state and the goal state of the server, and reports these metrics to the Azure Fabric Controller. If a fault state is reported by the Agent, the Fabric Controller can manage a reboot of the server or undertake re-provisioning of running application services from the current server to other healthy servers. Besides managing servers and load balancers, the Fabric Controller is also in charge of service provisioning by supporting a declarative service model. Declarative service specifications is encoded in every application service, which is used by the Fabric Controller for matching the services/VMs that meet required demands of CPU, bandwidth, operating system, redundancy tolerance and etc. However, as of the date this book chapter is being written (Oct 2009), most of these configurations are not available. However we expect Microsoft to release a lot of these roadmap features in Nov 2009 at PDC (Microsoft Professional Developer conference).

GoGrid Cloud Hosting offers developers up to three F5 Load Balancers for each account for distributing application service traffic across servers, as long as IPs and specific ports of these servers are attached into the load balancers. The load balancer implements two algorithms for routing application service requests. Round Robin algorithm distributes the incoming traffic to servers in sequence, one after another by taking turns in a distributed fashion. And Least Connect algorithm keeps routing incoming messages to the server that maintains least connection/request sessions. If the load balancer detects that a server crash has happened then all future requests will be bypassed from the crashed server, and would be redirected to other available servers. Currently, GoGrid Cloud Hosting only gives developers a programmatic API to implement their custom auto-scaling service. This is in contrast with the Amazon EC2 and Azure Cloud platforms that offer fully functional application scaling and load-balancing services. GoGrid Developers have to write a piece of code to collect usage data, and run/stop servers or migrate up to other servers based on collected data themselves.

Unlike other Cloud platforms, Google App Engine offers developers a scalable platform in which applications can run, rather than providing access directly to a customized virtual machine. Therefore, access to the underlying operating system is restricted in App Engine. And load-balancing strategies, service provisioning and auto scaling are all auto-magically managed by the system behind the scenes.

In addition, no single Cloud infrastructure provider has their data centers at all possible locations throughout the world. As a result Cloud application service (SaaS) providers will have difficulty in meeting QoS expectations for all their users. Hence, they would like to logically construct hybrid Cloud infrastructures (mixing multiple public and private clouds) to provide better support for their specific user needs. This kind of requirements often arises in enterprises with global operations and applications such as Internet service, media hosting, and Web 2.0 applications. This necessitates building technologies and algorithms for seamless integration of Cloud infrastructure service providers for provisioning of services across different Cloud providers.

# 6 CLOUD SERVICE DISCOVERY AND LOAD-BALANCING USING DHT OVERLAY

### *6.1 Distributed Hash Tables*

Structured systems such as DHTs offer deterministic query search results within logarithmic bounds on network message complexity. Peers in DHTs such as Chord, CAN, Pastry and Tapestry maintain an index for *O (log n)* peers where n is the total number of peers in the system. Inherent to the design of a DHT are the following issues [11]: (i) generation of node-ids and object-ids, called keys, using cryptographic/randomizing hash functions such as SHA-1 [15][17][30]. The objects and nodes are mapped on the overlay network depending on their key value. Each node is assigned responsibility for managing a small number of objects; (ii) building up routing information (routing tables) at various nodes in the network. Each node maintains the network location information of a few other nodes in the network; and (iii) an efficient look-up query resolution scheme.

Whenever a node in the overlay receives a look-up request, it must be able to resolve it within acceptable bounds such as in *O (log n)* routing hops. This is achieved by routing the look-up request to the nodes in the network that are most likely to store the information about the desired object. Such probable nodes are identified by using the routing table entries. Though at the core various DHTs (Chord, CAN, Pastry, and Tapestry etc.) are similar, still there exist substantial differences in the actual implementation of algorithms including the overlay network construction (network graph structure), routing table maintenance and node join/leave handling. The performance metrics for evaluating a DHT include fault-tolerance, load-balancing, efficiency of lookups and inserts and proximity awareness [31]. In Table-2, we present the comparative analysis of Chord, Pastry, CAN and Tapestry based on basic performance and organization parameters. Comprehensive details about the performance of some common DHTs under churn can be found in [32].

| DHT System | Overlay Structure | Lookup Protocol | Network Parameters | Routing Table Size | Routing Complexity | Join/Leave Overhead |
|---|---|---|---|---|---|---|
| Chord | Circular identifier space | matching key and server-id | *n*=number of servers | $O(\log n)$ | $O(\log n)$ | $O((\log n)^2)$ |
| Pastry | Plaxton style mesh | matching key and prefix in server-id | *n*=number of servers in the network, *b*= base of the identifier | $O(\log_b n)$ | $O(b \log_b n) + b$ | $O(\log n)$ |
| CAN | multi-dimensional space | key, value pair map to a point in space | *n*=number of servers in the network, *d*=dimensions | $O(2d)$ | $O(d n^{1/d})$ | $O(2d)$ |
| Tapestry | plaxton style mesh | Matching suffix in server-id | *n*=number of servers in the network, *b*= base of the identifier | $O(\log_b n)$ | $O(b \log_b n) + b$ | $O(\log n)$ |

Table 2: Summary of Complexity of Distributed Hash Table Overlays.

Other classes of structured peer-to-peer systems such as Mercury [26] do not apply randomizing hash functions for organizing data items and nodes. The Mercury system organizes nodes into a circular overlay and places data contiguously on this ring. As Mercury does not apply hash functions, data partitioning among nodes is non-uniform. Hence it requires an explicit load-balancing scheme. In recent developments, new generation P2P systems have evolved to combine both unstructured and structured P2P networks. We refer to this class of systems as hybrid. Structella [22] is one such P2P system that replaces the random graph model of an unstructured overlay (Gnutella) with a structured overlay, while still adopting the search and content placement mechanism of unstructured overlays to support complex queries. Other hybrid P2P design includes Kelips [34] and its variants. Nodes in Kelips overlay periodically gossip to discover new members of the network, and during this process nodes may also learn about other nodes as a result of lookup communication. Other variant of Kelips allows routing table entries to store information for every other node in the system. However, this approach is based on assumption that system experiences low churn rate [32]. Gossiping and one-hop routing approach has been used for maintaining the routing overlay in the work [33].

### *6.2 Designing Complex Services over DHTs*

#### *6.2.1 Limitations of Basic DHT Implementations & Query Types*

Traditionally, DHTs have been efficient for *single-dimensional* queries such as “finding all resources that match the given attribute value”. Since Cloud computing IaaS and PaaS level services such as servers, VMs, enterprise computers (private cloud resources), storage devices, and databases are identified by more than one attribute; therefore a search query for these services is always *multi-dimensional*. These search dimensions or attributes can include service type, processor speed, architecture, installed operating system, available memory, and network bandwidth.

Based on recent information published by Amazon EC2 CloudWatch service, each Amazon Machine Image (AMI) instance has seven performance metrics (see Table-3) and three dimensions (see Table-4) associated with it. Additionally, these AMIs can host different application service types including web hosting, social networking, content-delivery, and high-performance computing that have varying request invocation, access and distribution pattern. The type of application services hosted by an AMI instance is dependent on the business needs and scientific experiments. In these cases a *Cloud service discovery query* (which can be issued by provisioning software) will combine the aforementioned attributes related to AMI instances and application service types and therefore can have the following semantics:

*Cloud Service Type = “web hosting” && Host CPU Utilization < “50%” && Instance OSType = “WinSrv2003” && Host Processor Cores > “1” && Host Processors Speed > “1.5 GHz” && Host Cloud Location = “Europe”*

| **CPU Utilization** | **Network Incoming Traffic** | **Network Outgoing Traffic** | **Disk Write Operations** | **Disk Read Operations** | **Disk Write Bytes** | **Disk Read Bytes** |
|---|---|---|---|---|---|---|

Table 3: Performance metrics associated with an Amazon EC2 AMI instance.

| Image ID | Auto Scaling Group Name | Instance ID | Instance Type |
|---|---|---|---|

Table 4: Performance dimensions associated with an Amazon EC2 AMI instance.

On the other hand, VM instances deployed on the Cloud hosts needs to publish their information so that provisioning software can search and discover them. VM instances update their software and hardware configuration and the deployed services' availability status by sending *update query* to the DHT overlay. The service configuration distribution in three dimensions is shown in Fig. 3. An update query has the following semantics:

*Cloud Service Type = "web hosting" && Host CPU Utilization = "30%" && Instance OSType = "WinSrv2003" && Host Processor Cores = "2" && Host Processors Speed = "1.5 GHz" && Host Cloud Location = "Europe"*

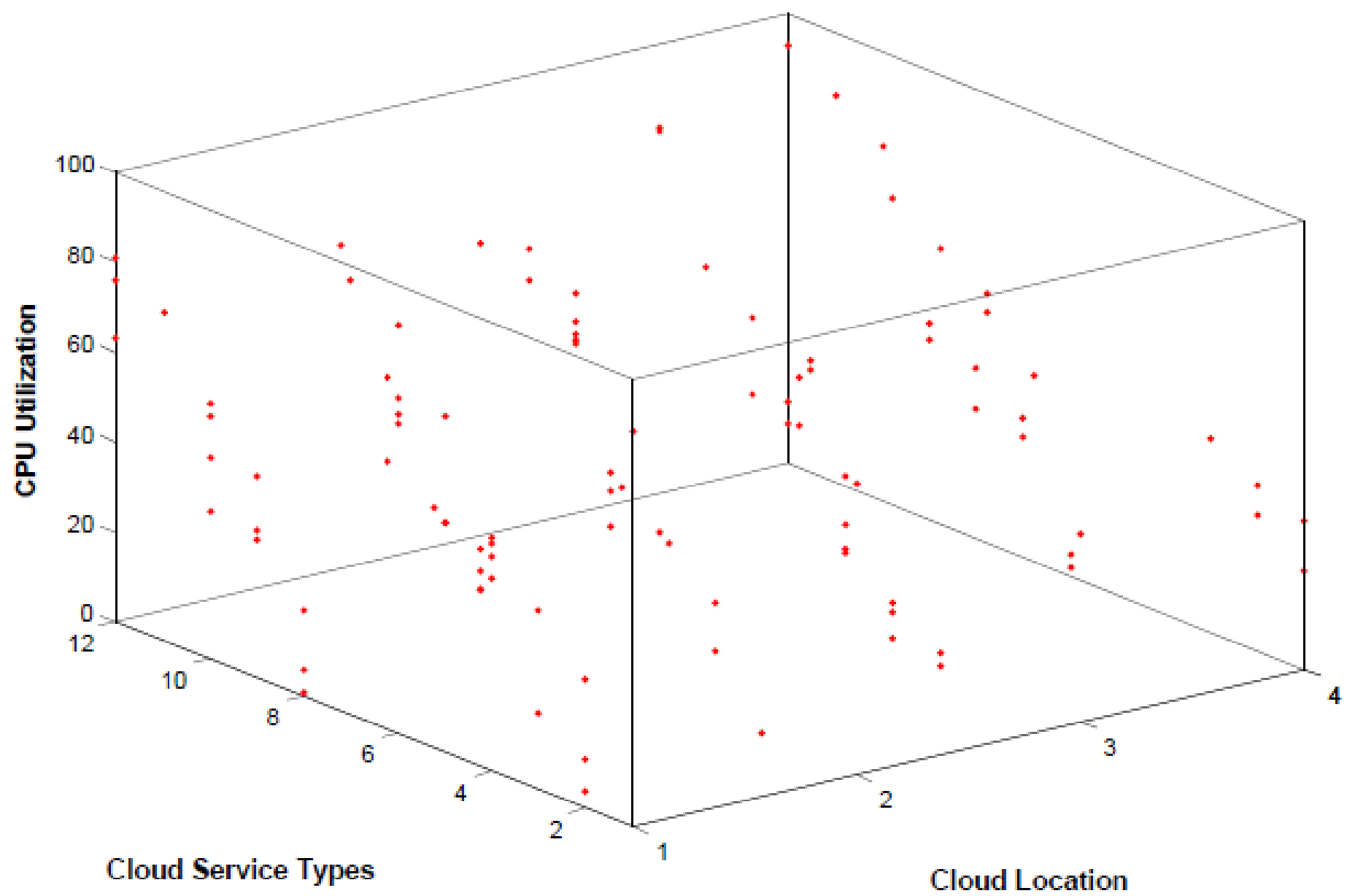

Figure 3: Service configuration and update query distribution in 3-D space.

Extending DHTs to support indexing and matching of multi-dimensional range (service discovery query) or point (update query) queries, to index all resources whose attribute value overlap a given search space, is a complex problem. Multi-dimensional range queries are based on ranges of values for attributes rather than on specific values. Compared to single-dimensional queries, resolving multi-dimensional queries is far more complicated, as there is no obvious total ordering of the points in the attribute space. Further, the query interval has varying size, aspect ratio and position such as a window query. The main challenges involved in enabling multi-dimensional queries in a DHT overlay include designing efficient service attribute data: (i) distribution or indexing techniques; and (ii) query routing techniques.

*6.2.2 Data Indexing Techniques for Mapping Multi-dimensional Range and Point Queries*

A data indexing technique partitions the multi-dimensional attribute space over the set of VMs in a DHT network. Efficiency of the distribution mechanism directly governs how the query processing load is distributed among the Cloud peers. A good distribution mechanism should possess the following characteristics [36]: (i) locality: data points nearby in the attribute space should be mapped to the same Cloud peer, hence limiting the distributed lookup complexity; (ii) load balance: the number of data points indexed by each Cloud peer should be approximately the same to ensure uniform distribution of query processing; (iii) minimal metadata: prior information required for mapping the attribute space to the overlay space should be minimal; and (iv) minimal management overhead: during VM instantiation and destruction operation, update policies such as the transfer of data points to a newly joined Cloud peer should cause minimal network traffic. Note that, assumption here is every VM instance hosts a Cloud peer service, which is responsible for managing activities related to overlay network.

There are different kinds of database indices [21] that can handle mapping of multi-dimensional objects such as the Space Filling Curves (SFCs) (including the Hilbert curves, Z-curves), k-d tree, MX-CIF Quad tree and R*-tree in a DHT overlay. In literature these indices are referred to as *spatial indices* [35]. Spatial indices are well suited for handling the complexity of multi-dimensional queries. Although some spatial indices can have issues as regards to routing load-balance in case of a skewed attribute/data set, all the spatial indices are generally scalable in terms of the number of hops traversed and messages generated while searching and routing multi-dimensional/spatial service discovery and update queries. However, there are different tradeoffs involved with each of the spatial indices, but basically they can all support scalability and Cloud service discovery. Some spatial index would perform optimally in one scenario but the performance could degrade if the attribute/data distribution changed significantly.

### *6.2.3 Routing Techniques for Handling Multi-dimensional Queries in DHT overlay*

DHTs guarantee deterministic query lookup with logarithmic bounds on network message cost for single-dimensional queries. However, Cloud's service discovery and update query are multi-dimensional (as discussed in previous sections). Hence, existing DHT routing techniques need to be augmented in order to efficiently resolve multi-dimensional queries. Various data structures that we discussed in previous section effectively create a logical multi-dimensional index space over a DHT overlay. A look-up operation involves searching for an index or set of indexes in a multi-dimensional space. However, the exact query routing path in the multi-dimensional logical space is directly governed by the data distribution mechanism (i.e. based on the data structure that maintains the indexes). In this context, various approaches have proposed different routing/indexing heuristics.

Efficient query routing algorithm should exhibit the following characteristics [36]: (i) routing load balance: every peer in the network on the average should route forward/route approximately same number of query messages; and (ii) low routing state per Cloud peer: each Cloud peer should maintain a small number of routing links hence limiting new Cloud peer (VM) join and Cloud peer (VM) state update cost. In the current peer-to-peer literature, multi-dimensional data distribution mechanisms based on the following structures have been proposed: (i) space filling curves; and (ii) tree-based structures. Resolving multi-dimensional queries over a DHT overlay that utilizes SFCs for data distribution consists of two basic steps [37]: (i) mapping the multi-dimensional query onto the set of relevant clusters of SFC-based index space; and (ii) routing the message to all VMs that fall under the computed SFC-based index space. On the other hand, routing multi-dimensional query in a DHT overlay that employ tree-based structures for data distribution requires routing to start from the root. However, the root VM presents a single point of failure and load imbalance. To overcome this, the authors in introduced the concept of fundamental minimum level. This means that all the query processing and the data storage should start at that minimal level of the tree rather than at the root. There are number of techniques available for distributed routing in multi-dimensional space. The performance of techniques varies depending upon the distribution of data in the multi-dimensional space, and VM in the underlying DHT overlay.

## 5 CLOUD PEER SOFTWARE FABRIC: DESIGN AND IMPLEMENTATION

The Cloud peer implements services for enabling decentralized and distributed discovery supporting status lookups and updates across the inter-networked Cloud computing systems; enabling inter-application service coordinated provisioning for optimizing load-balancing and tackling the distributed service contention problem. Dotted box in Fig. 2 shows the layered design of Cloud peer service over DHT based self-organizing routing structure. The services build upon the DHT routing structure extends (both algorithmically and programmatically) the fundamental properties related to DHTs including deterministic lookup, scalable routing, and decentralized network management. The Cloud peer service is divided into a number of sub-layers (see Fig. 2): (i) higher level services for discovery, coordination, and messaging; (ii) low level distributed indexing and data organization techniques, replication algorithms, and query load-balancing techniques; (iii) DHT-based self-organizing routing structure. A Cloud peer undertakes the following critical tasks that are important for proper functioning of DHT-based provisioning overlay:

### *5.1 Overlay Construction*

The overlay construction refers to how Cloud peers are logically connected over the physical network. The software implementation utilizes (the open source implementation of Pastry DHT known as the FreePastry) Pastry [27] as the basis for creation of Cloud peer overlay. A Pastry overlay inter-connects the Cloud peer services based on a ring topology. Inherent to the construction of a Pastry overlay are the following issues: (i) Generation of Cloud peer ids and query (discovery, update) ids, called keys, using cryptographic/randomizing hash functions such as SHA-1. These ids are generated from from 160-bit unique identifier space. The id is used to indicate a Cloud peer's position in a circular id space, which ranges from $0$ to $2^{160}-1$. The queries and Cloud peers are mapped on the overlay network depending on their key values. Each Cloud peer is assigned responsibility for managing a small number of queries; and (ii) Building up routing information (leaf set, routing table, and neighborhood set) at various Cloud peers in the network. Given the Key K, Pastry routing algorithm can find the Cloud peer responsible for this key in $O(\log_b n)$ messages, where $b$ is the base and $n$ is the number of Cloud Peers in the network.

Each Cloud peer in the Pastry overlay maintains a routing table, leaf set, and neighborhood set. These tables are constructed when a Cloud peer joins the overlay, and it is periodically updated to take into account any new joins, leaves or failures. Each entry in the routing table contains the IP address of one of the potentially many Cloud peers whose id have the appropriate prefix; in practice, a Cloud peer is chosen, which is close to the current peer, according the proximity metric. Fig. 4 shows a hypothetical Pastry overlay with keys and Cloud peers distributed on the circular ring based on their cryptographically generated ids.

### *5.2 Multi-dimensional Query Indexing*

In order to support multi-dimensional query indexing (Cloud service type, Host utilization, Instance OS type, Host Cloud location, Host Processor speed) over Pastry overlay, a Cloud peer implements a distributed indexing technique [16], which is a variant of peer-to-peer MX-CIF Quad tree [38] data structure. The distributed index builds a multi-dimensional attribute space based on the Cloud service attributes, where each attribute represents a single dimension. An example 2-dimensional attribute space that indexes service attributes including Speed and Cpu Type is shown in Fig. 4. First step in initializing the distributed index is the process called Minimum Division ($f_{min}$). This process divides the Cartesian space into multiple index cells when the multi-dimensional distributed index is first created. As a result of this process, the attribute space resembles a grid like structure consisting of multiple index cells. The cells resulting from this process remain constant throughout the life of the indexing domain and serve as entry points for subsequent service discovery and update query mapping. The number of cells produced at the minimum division level is always equal to $(f_{min})^{dim}$, where *dim* is dimensionality of the attribute space. Every Cloud peer in the network has basic information about the attribute space coordinate values, dimensions and minimum division level. Cloud peers can obtain this information (cells at minimum division level, control points) in a configuration file from the bootstrap peer. Each index cell at $f_{min}$ is uniquely identified by its centroid, termed as the *control point*. In Fig. 4, $f_{min}$ = 1, dim=2. The Pastry overlay hashing method (DHash(coordinates)) is used to map these control points so that the responsibility for an index cell is associated with a Cloud peer in the overlay. For example in Fig.4, DHash(*x1, y1*) = *k10* is the location of the control point A (*x1,y1*) on the overlay, which is managed by Cloud peer *12*.

### *5.3 Multi-dimensional Query Routing*

This action involves the identification of the index cells at minimum division level $f_{min}$ in the attribute space to map a service discovery and update query. For mapping service discovery query, the mapping strategy depends on whether it is a multi-dimensional point query (equality constraints on all search attribute values) or multi-dimensional range query. For a multi-dimensional point service discovery query the mapping is straight forward since every point is mapped to only one cell in the attribute space. For a multi-dimensional, mapping is not always singular because a range look-up can cross more than one index cell. To avoid mapping a multi-dimensional service discovery query to all the cells that it crosses (which can create many unnecessary duplicates) a mapping strategy based on *diagonal hyperplane* of the attribute space is utilized. This mapping involves feeding the service discovery query's spatial dimensions into a mapping function, *Imap*(query). This function returns the IDs of index cells to which given query can be mapped~(refer to step 7 in Fig. 4. Distributed hashing (*DHash*(cells)) is performed on these IDs (which returns keys for Pastry overlay) to identify the current Cloud peers responsible for managing the given keys. A Cloud peer service uses the index cell(s) currently assigned to it and a set of known base index cells obtained at the initialisation as the candidate index cells. Similarly, mapping of update query also involves the identification of the cell in the attribute space using the same algorithm. A update query is always associated with an event region [39] and all cells that fall fully or partially within the event region would be selected to receive the corresponding objects. The calculation of an event region is also based upon the diagonal hyperplane of the attribute space. Giving in depth information here is out of the scope for this chapter, however the readers who would like to have more information can refer the paper [16][20][21] that describes the index in detail.

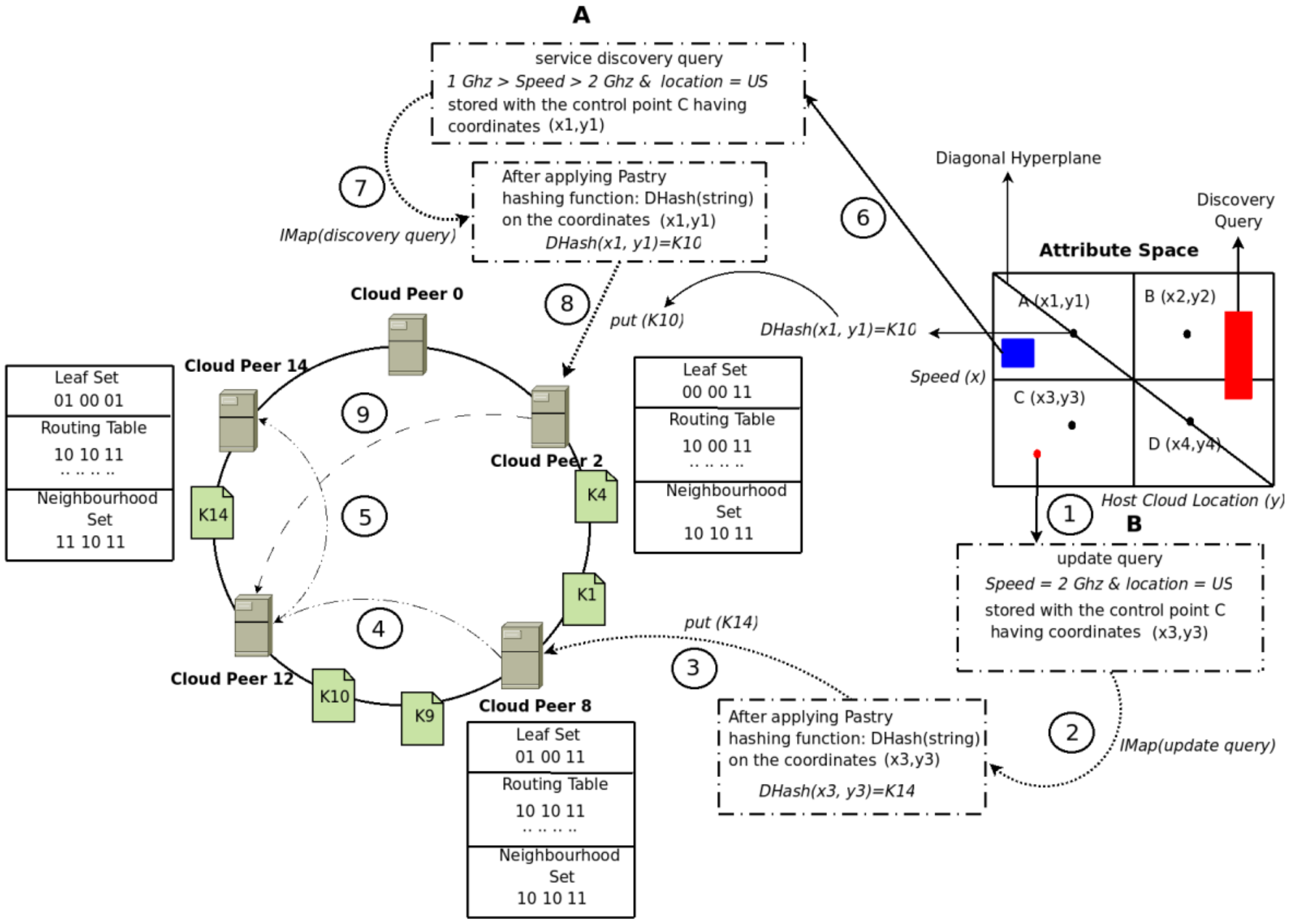

Figure 4: A pictorial representation of Pastry (DHT) overlay construction, multi-dimensional data indexing, and routing: (1) a service hosted within a VM publishes a update query; (2) Cloud peer 8 computes the index cell, C(x3,y3), to which the update query maps by using mapping function *IMap*(query); (3) Next, distributed hashing function, *DHash*(x3, y3), is applied on the cell's coordinate values, which yields a overlay key, K14; (4) Cloud peer 8 based on its routing table entry forwards the request to peer 12; (5) Similarly, peer 12 on the overlay forwards the request to Cloud peer 14; (6) a provisioning service submits a service discovery query; (7) Cloud peer 2 computes the index cell, C(x1, y1), to which the service discovery query maps; (8) *DHash*(x1, y1) is applied that yields an overlay key, K10; (9) Cloud peer 2 based on its routing table entry forwards the mapping request to peer 12.

### *5.4 Designing Decentralized and Coordinated Load-Balancing Mechanism*

A coordinated provisioning of requests between virtual machine instances deployed in Clouds is critical, as it prevents the service provisioners from congesting the particular set of VMs and network links, which arises due to lack of complete global knowledge. In addition, it significantly improves the Cloud user Quality of Service (QoS) satisfaction in terms of response time. The Cloud peer service in conjunction with the Pastry overlay and multi-dimensional indexing technique is able to perform a decentralized and coordinated balancing of service provisioning requests among the set of available VMs. The description of the actual load-balancing mechanism follows next.

As mentioned in previous section, both service discovery (issued by service provisioner) and update query (published by VMs or Services hosted within VMs) queries are spatially hashed to an index cell *i* in the multi-dimensional attribute space. In Figure 5, service discovery query for provisioning request P1 is mapped to index cell with control point value A, while for P2, P3, and P4, the responsible cell has control point value C. Note that, these service discovery queries are posted by service provisioners. In Figure 4, a provisioning service inserts a service discovery query with Cloud peer p, which is mapped to index cell i. The index cell i is spatially hashed through IMap(query) function to an Cloud peer s. In this case, Cloud peer s is responsible for coordinating the provisioning of services among all the service discovery queries that are currently mapped to the cell i. Subsequently, VM u issues a resource ticket (see Figure 5) that falls under a region of the attribute space currently required by the provisioning requests P3 and P4. Next, the Cloud peer s has to decide which of the requests (either P3 or P4 or both) is allowed to claim the update query published by VM u. The load-balancing decision is based on the principle that it should not lead to over-provisioning of service(s) hosted within VM u. This mechanism leads to coordinated load-balancing across VMs in Clouds and aids in achieving system-wide objective function.

| **Time** | **Discovery Query ID** | **Service Type** | **Speed (GHz)** | **Cores** | Location |
|---|---|---|---|---|---|
| 300 | Query 1 | Web Hosting | > 2 | 1 | USA |
| 400 | Query 2 | Scientific Simulation | > 2 | 1 | Singapore |
| 500 | Query 3 | Credit Card Authenticator | > 2.4 | 1 | Europe |

Table 5: Service discovery query stored with a Cloud Peer service at time *T.*

| **Time** | **VM ID** | **Service Type** | **Speed (GHz)** | **Processors** | Type |
|---|---|---|---|---|---|
| 700 | VM 2 | Credit Card Authenticator | 2.7 | 1 (available) | Europe |

Table 6: Update query published with a Cloud Peer service at time *T.*

The examples in Table 5 are list of service discovery queries that are stored with a Cloud peer service at time T = 700 secs. Essentially, the queries in the list arrived at a time <= 700 and waited for a suitable update query that can meet their provisioning requirements (software, hardware, service type, location). Table 6 depicts an update query that has arrived at T = 700. Following the update query arrival, the Cloud peer service undertakes a procedure that allocates the available service capacity with VM (that published the update query) among the list of matching service discovery queries. Based on the updating VM's attribute specification, only service discovery query 3 matches. Following this, the Cloud peer notifies the provisioning services that posted the Query 3. Note that Query 1 and 2 have to wait for the arrival of update queries that can match their requirements.

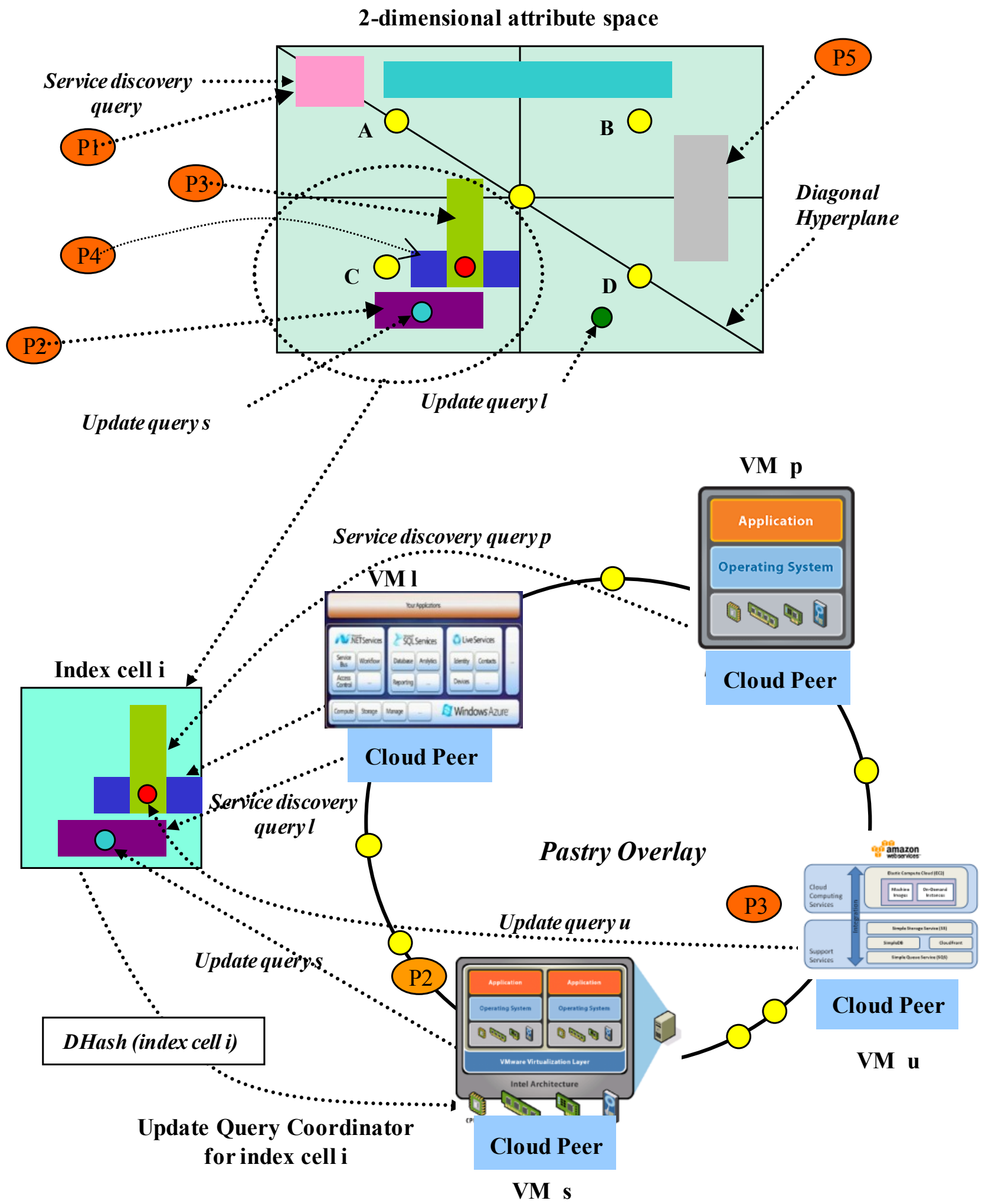

Figure 5: Coordinated provisioning across VM instances. multi-dimensional service provisioning requests {P1, P2, P3, P4}, index cell control points {A, B, C, D}, multi-dimensional update queries {l, s} and some of the spatial hashings to the Pastry overlay, i.e. the *multi*-dimensional (spatial) coordinate values of a cell's control point is used as the Pastry key. For this Figure $f_{min}$ =2, dim = 2.

## 6. Experiments and Evaluation

In this section, we evaluate the performance of the proposed peer-to-peer Cloud provisioning concept by creating a service and VM pool that consists of multiple virtual machines that are hosted within the Amazon EC2 infrastructure. We assume unsaturated server availability for these experiments, so that enough capacity can always be allocated to a VM for any service request. Next, we describe the various details related to Cloud Peer (peer-to-peer network, multi-dimensional index structure, and network configuration parameters), PaaS layer provisioning software, and application characteristics related to this experimental evaluation.

### *6.1 Cloud Peer Details*

A Cloud peer service operates at PaaS layer and handles activities related to decentralized query (discovery and update) routing, management, and matching. Additionally, it also implements the higher level services such as publish/subscribe based coordinated interactions and service selections. Every VM instance, which is deployed on a Cloud platform, hosts a Cloud peer service (see Fig. 4 and Fig. 5) that loosely glues it to the overlay. Next, follows the details related to Cloud peer configuration.

*FreePastry Network Configuration*: Both Cloud Peers' nodeIds and discovery/update queries' IDs are randomly generated from and uniformly distributed in the 160-bit Pastry identifier space. Every Cloud peer service is configured to buffer maximum of 1000 messages at a given instance of time. The buffer size is chosen to be sufficiently large such that the FreePastry[1] does not drop any messages. Other network parameters are configured to the default values as given in the file freepastry.params. This file is provided with the FreePastry distribution.

*Multi-dimensional Index Configuration*: The minimum division $f_{min}$ of logical *multi*-dimensional index that forms the basis for mapping, routing, and searching the service discovery and update queries is set to 3, while the maximum height of the distributed index tree, $f_{max}$ is constrained to 3. In other words, the division of the *multi*-dimensional attribute space is not allowed beyond $f_{min.}$ This is done for simplicity, understanding the load balancing issues of multi-dimensional indices with increasing $f_{max}$ is a different research problem and is beyond scope of this chapter. The index space has provision for defining service discovery and update queries that specify the VM characteristics in *4* dimensions including number of application service type being hosted, number of processing cores available on the server hosting the VM, hardware architecture of the processor(s), and their processing speed. The aforementioned multi-dimensional index configuration results into $81(3^4)$ index cells at $f_{min}$ level.

*Service Discovery and Update Query's Multi-dimensional Extent*: Update queries, which are posted by VM instances, express equality constraints on service, installed software environments and hardware configuration attribute values (e.g. =). In other words, update queries are always *multi*-dimensional point query for this study. On the other hand, the service discovery queries posted by Application Provisioner have their extent in multiple dimensions with both, range and fixed constraint (e.g. >=, <=) for the attributes. The spatial extent of a discovery query in different attribute dimension is controlled by the characteristic of the application services VM, hosting node, and workload patterns. Attributes including application service type, processor architecture, and number of processors are fixed, i.e. they are expressed as equality constraints. The value for processing speed is expressed using >= constraints, i.e. search for the services, which can process application atleast as fast as given performance threshold. Deriving performance threshold is a complicated research problem which depends on application workload characteristics, infrastructure performance history, and quality of service requirements.

[1] FreePastry: An Open Source Pastry DHT Implementation. http://freepastry.rice.edu/FreePastry.

### *6.2 Aneka: PaaS Layer Application Provisioning and Management Service*

At PaaS layer, we utilize the Aneka [23] software framework that handles activities related to application element scheduling, execution, and management. Aneka is a .NET-based service oriented platform for constructing Cloud computing environments. To create a Cloud application provisioning systems using Aneka, a developer or application scientist needs to start an instance of the configurable Aneka container hosting required services on each selected VMs.

Services of Aneka can be clearly characterized into two distinct spaces: (i) Application Provisioner: This service implements the functionality that accepts application workload from Cloud users, performs dynamic discovery of application management services via the Cloud peer service, dispatches workload to application management service, monitors the progress of their execution, and collects the output data, which returned back to the Cloud users. A Application Provisioner need not be hosted within a VM, it only needs to know the end-point address (such as web service address) of a random Cloud peer service in the overlay to which it can connect and submit its service discovery query; and (ii) Application Management Service: This service, which is hosted within a VM, is responsible for handling execution and management of submitted application workload. An application management service sits within a VM and updates its usage status, software, and hardware configurations by sending update queries to the overlay. One or more instance of application management service can be connected in a single-level hierarchy to be controlled by a root level Aneka Management Coordinator. This kind of service integration is aimed at making application programming flexible, efficient, and scalable.

An interesting point to note here is that Cloud peer service is completely detached from the application provisioning and application management service (Aneka). Hence, this presents an opportunity for developers and scientist to develop application specific provisioning and management software that can transparently connect to the Cloud peer service through standard WS* APIs.

### *6.3 Test Application*

The PaaS layer software service, Aneka, supports composition and execution of application programs that are composed using different service models to be executed within the same software environment. The experimental evaluation in this chapter considers execution of applications programmed using multi-threaded programming model. The Thread programming model [23] defines an application as a collection of one or more independent work units. This model can be successfully utilized to compose and program embarrassingly parallel programs (parameter sweep applications). The Thread model fits better for implementing and architecting new applications, algorithms on Cloud infrastructures since it gives finer degree of control and flexibility as regards to runtime control.

To demonstrate the feasibility of architecting Cloud provisioning services based on peer-to-peer network models, we consider composition and execution of Mandelbrot Set computation. Mathematically, the Mandelbrot set is an ordered collection of points in the complex plane, the boundary of which forms a fractal. The Application Provisioner service implements and cloud enables the Mandelbrot fractal calculation using a multi-threaded programming model. The application submission interface allows the user to configure number of horizontal and vertical partitions into which the fractal computation can be divided. The number of independent thread units created is equal to the horizontal x vertical partitions. For evaluations, we vary the values for horizontal and vertical parameters over the interval [5 X 5, 10 X 10, and 15 X 15]. This configuration results in observation points.

### *6.4 Deployment of Test Services on Amazon EC2 Platform*

In order to test the feasibility of aforementioned services in regards to the provisioning of application services on Amazon EC2 cloud platform, we created Amazon Machine Images (AMIs) packaged with a Cloud peer, Application Management and Aneka Management Coordinator services. The image that hosts the Aneka Management Coordinator is equipped with Microsoft Windows Server 2003 R2 Datacenter edition, Microsoft SQL Server 2005 Express and Internet Information Services 6, while the AMI hosts only the Management Service has Microsoft Windows Server 2003 R2 Datacenter system installed. For every AMI, we installed only the essential software including mandatory Cloud peer service, which is hosted within a Tomcat 6.0.10, Axis2 1.2 container. Cloud peer is exposed to the provisioning and management services through WS* interfaces. Later, we build our customized Amazon Machine Images from the two instances, creating and starting up more Management Coordinator and Application Management services by using customized images. We configured three Management Coordinators and nine Management Services. The Management Service is divided into groups of three that connect with a single Coordinator resulting in a hierarchical structure. The Management Coordinator services communicate and inter-network through the Cloud peer fabric service. Fig. 6 shows the pictorial representation of the experiment setup.

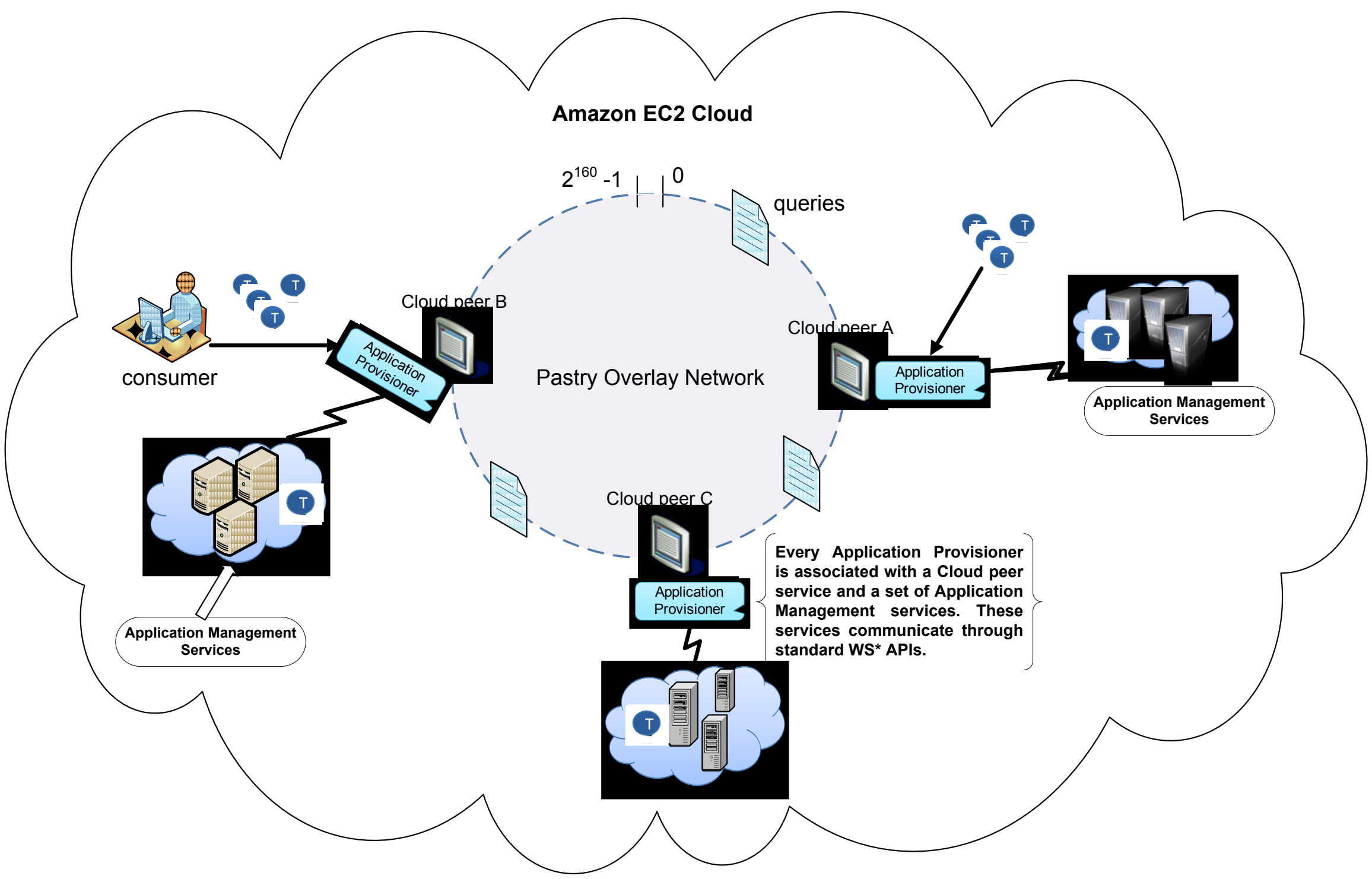

Figure 6: Experiment Setup in Amazon EC2.

### *6.5 Results and Discussions*

To measure performance of peer-to-peer Cloud provisioning technique in regards to response time, coordination delay and Pastry overlay network message complexity; we consider simultaneous provisioning of test applications at Application Provisioner A and B (see Fig. 6). The *response time* for an application is calculated by subtracting the output arrival time of the last thread in the submission list from the time at which the application is submitted. The metric *coordination delay* sums up the latencies for: (i) a service discovery query to be mapped to a Cloud peer, (ii) waiting time till a update query matches the discovery query; and (ii) notification delay from the Cloud peer to the Application Provisioner that originally posted the service discovery query. Pastry overlay message complexity measures the details related number of messages that flow through the network in order to: (i) initialize the multi-dimensional attribute space, (ii) map the discovery and update queries, (iii) maintain overlay, and (iv) send notifications.

Fig. 7 shows the results for response time in seconds with increasing complexity/problem size for the test application. Cloud consumers submit their applications with Provisioner A and B. The initial experimental results reveal that with increase in problem complexity (number of horizontal and vertical partitions), the Cloud consumers experience increase in response times. The basic reason behind this behavior of the provisioning system is related to the fixed number Application Management services, i.e. 9, available to the Application Provisioners. With increase in the problem complexity, the number of job threads (a job thread represents a single work unit, e.g. For 5 X 5 Mandelbrot configuration, 25 job threads are created and submitted with the Application Provisioner) that are to be executed with Management services increase, hence leading to worsening queuing and waiting delays. However, this behavior of the provisioning system can be fixed through implementation of reactive provisioning of new VM instances to reflect the sudden surge in application workload processing demands (problem complexity). In our future work we want to explore how to dynamically provision or de-provision VMs and associated Application Management services driven by workload processing demands.

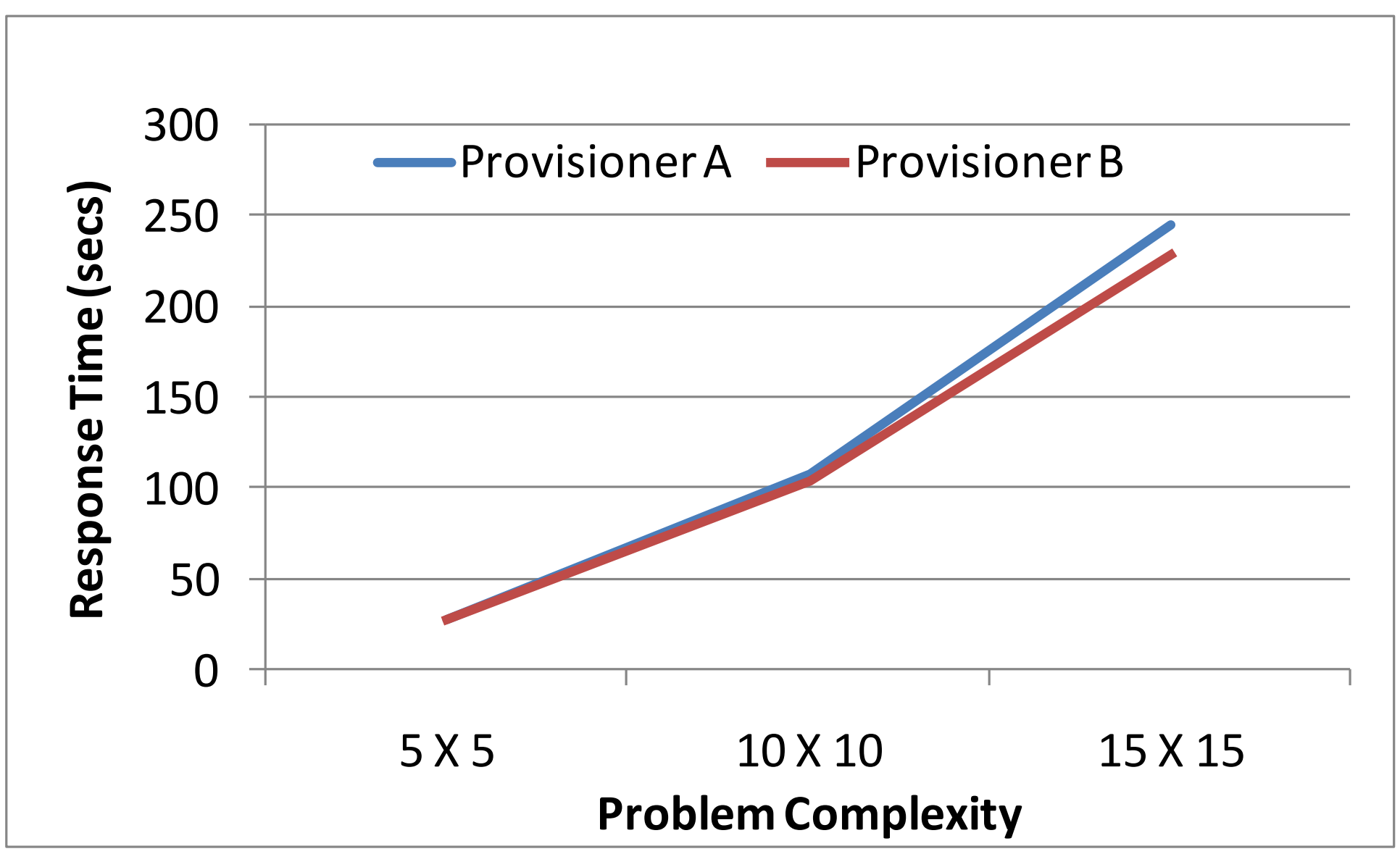

Figure 7: Response Time (secs) Vs. Problem Complexity.

Fig. 8 presents the measurements for average coordination delay for each discovery query with respect to increase in the problem complexity. The results show that at higher problem complexity, the discovery queries experience increased coordination delay. This happens due to the reason that the discovery queries of the corresponding job threads have to wait for longer period of time before they are matched against an update query object. However the job thread processing time (CPU time) is not affected by the coordination delay, hence the response time plot in Fig. 7 shows the similar trend to Fig. 8.

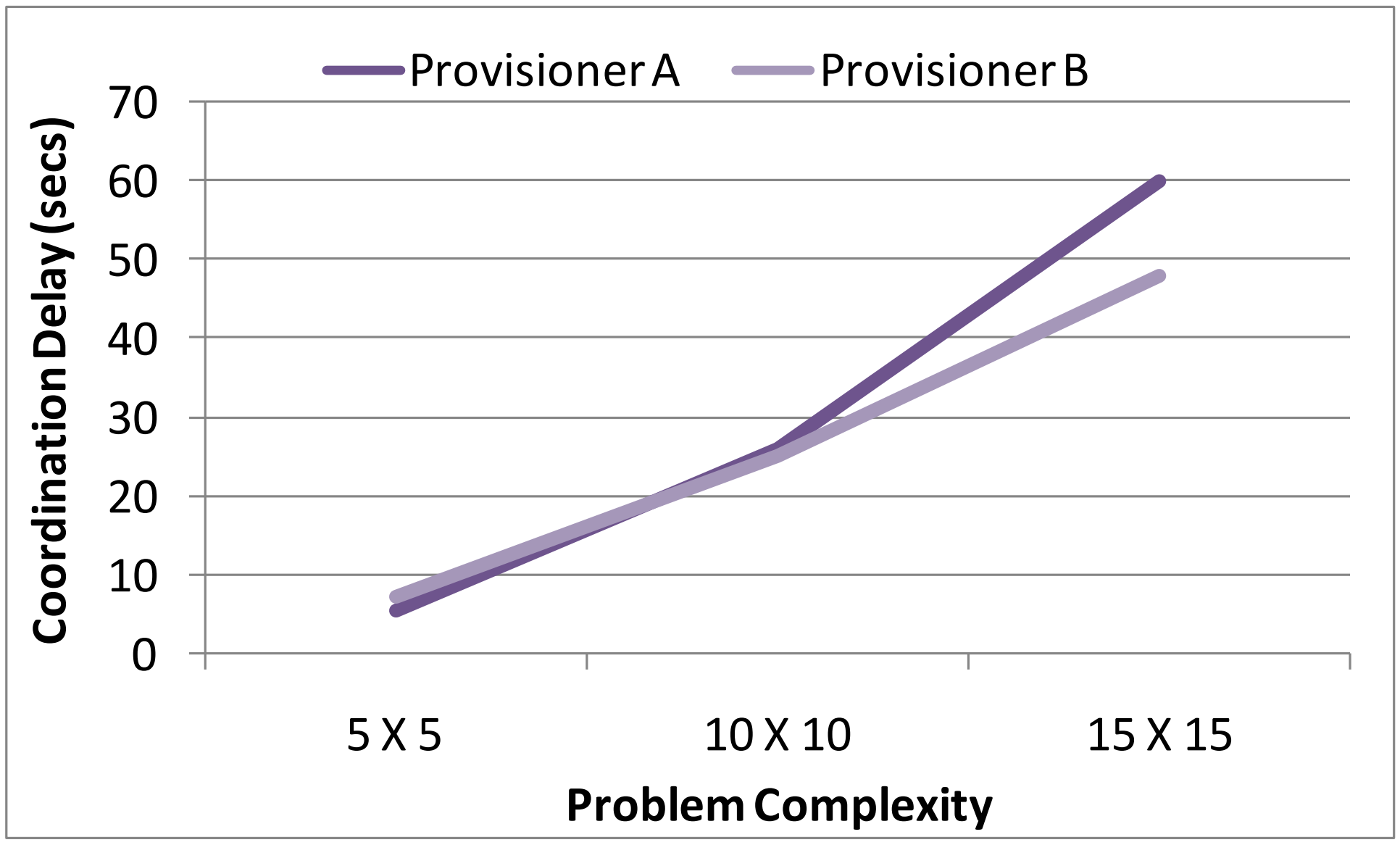

Figure 8: Coordination Delay (secs) Vs. Problem Complexity.

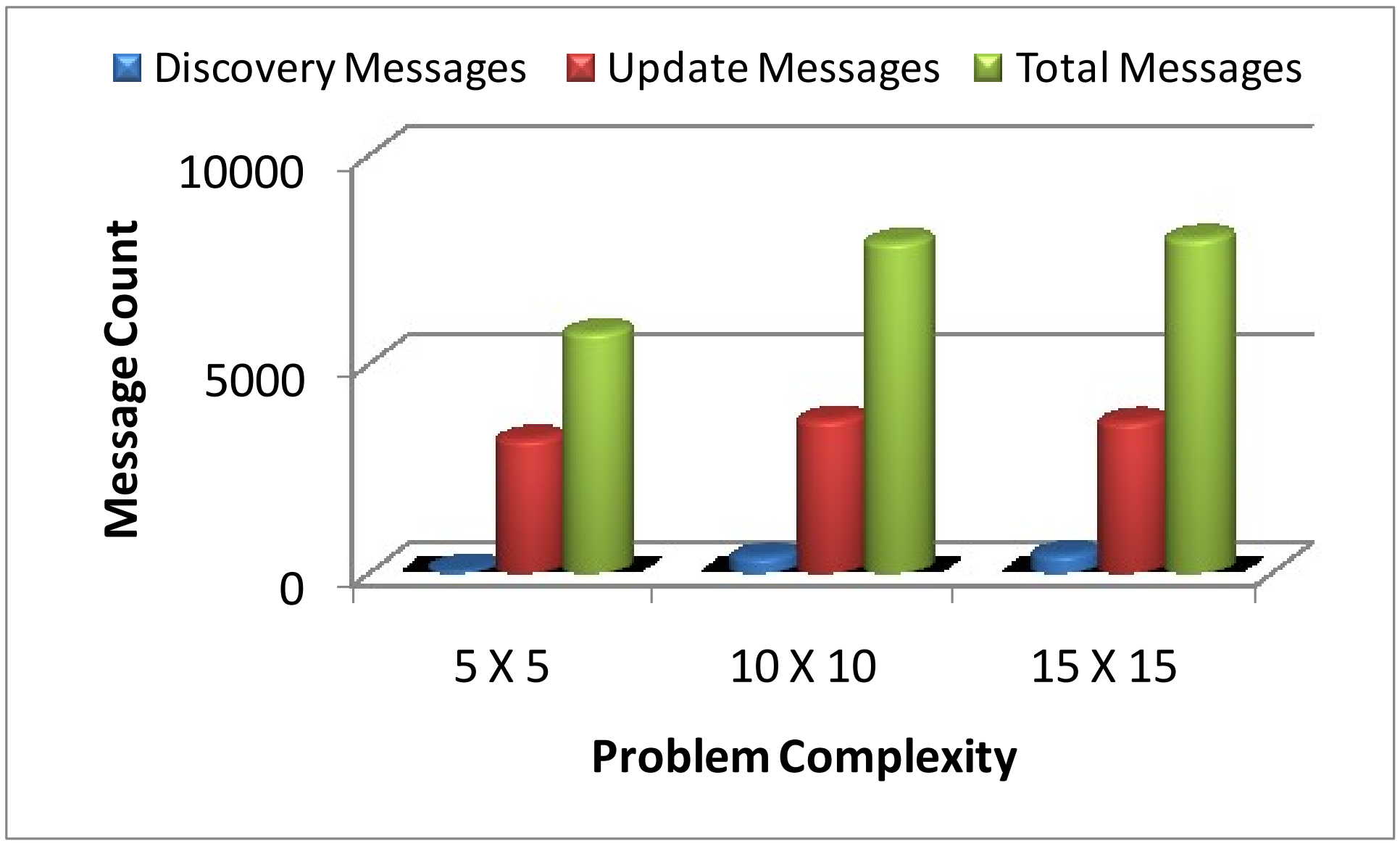

Figure 9: Message Count Vs. Problem Complexity.

In Fig. 9, we show the message overhead involved with management of multi-dimensional index, routing of discovery and update query messages, and maintenance of Pastry overlay. We can clearly see that as application size (problem complexity) increase the number of messages required for mapping the query objects and maintenance of the overlay network increase. The number of discovery and update messages produced in the overlay is a function of the multi-dimensional index structure that indexes and routes these queries in a distributed fashion. Hence, the choice of the multi-dimensional data indexing structure and routing technique governs the manageability and efficiency of the overlay network (latency, messaging overhead). Hence, there is much work required in this domain as regards to evaluating the performance of different types of multi-dimensional indexing structures for mapping the query messages in peer-to-peer settings.

## 7. Conclusions and Path Forward

Developing provisioning techniques that integrate application services in a peer-to-peer fashion is critical to exploiting the potential of Cloud computing platforms. Architecting provisioning techniques based on peer-to-peer network models (such as DHTs) is significant; Since peer-to-peer networks are highly scalable, can gracefully adapt to the dynamic system expansion (join) or contraction (leave, failure), are not susceptible to single point of failure. To this end, we presented a software fabric called *Cloud pee*r that creates an overlay network of VMs and application services for supporting scalable and self-managing service discovery and load-balancing. The functionality exposed by the Cloud peer service is very powerful and our experimental results conducted on Amazon EC2 platform confirms that it is possible to engineer and design peer-to-peer Cloud provisioning systems and techniques.

As part of our future work, we would explore other multi-dimensional data indexing and routing techniques that can achieve close to logarithmic bounds on messages and routing state, balance query (discovery, load-balancing, coordination) processing load, preserves data locality, and minimize the metadata. Another important algorithmic and programming challenge in building robust Cloud peer services is to guarantee consistent routing, lookup, and information consistency under concurrent leave, failure and join operations by application services. To address these issues, we will investigate robust fault-tolerance strategies based on distributed replication of attribute/query sub-spaces to achieve a high level of robustness and performance guarantees.